\documentclass[aps,preprint,nofootinbib,showpacs]{revtex4}
\usepackage{graphicx,color,amsmath,epsfig,rotating}
\newcommand{\be}{\begin{equation}}
\newcommand{\ee}{\end{equation}}
\newcommand{\ba}{\begin{eqnarray}}
\newcommand{\beq}{\begin{equation}}
\newcommand{\eeq}{\end{equation}}
\newcommand{\ea}{\end{eqnarray}}

\newcommand{\MNS}{{\text{MNS}}}

\def\beqa{\begin{eqnarray}}
\def\eeqa{\end{eqnarray}}

\def\bea{\begin{eqnarray}}
\def\eea{\end{eqnarray}}

\def\err#1#2{\lower2pt\hbox{ $\stackrel{\scriptstyle +#1}{\scriptstyle -#2}$}}
\def\ga{\mathrel{\raise.3ex\hbox{$>$\kern-.75em\lower1ex\hbox{$\sim$}}}}
\def\la{\mathrel{\raise.3ex\hbox{$<$\kern-.75em\lower1ex\hbox{$\sim$}}}}
\def\bmaT{\left(\begin{array}{ccc}}
\def\emaT{\end{array}\right)}
\def\bma{\left( \begin{array} }
\def\ema{\end{array} \right)}
\def\gsim{~{\rlap{\lower 3.5pt\hbox{$\mathchar\sim$}}\raise 1pt\hbox{$>$}}\,}
\def\lsim{~{\rlap{\lower 3.5pt\hbox{$\mathchar\sim$}}\raise 1pt\hbox{$<$}}\,}

\begin{document}

\title{\boldmath Doubly charged Higgs bosons and 
three-lepton \\ signatures in the Higgs Triplet Model\unboldmath} 

\author{A.G. Akeroyd}
\affiliation{Department of Physics and Center for Mathematics and Theoretical Physics, National Central University, Chungli, Taiwan 320, Taiwan}
\author{Cheng-Wei Chiang}
\affiliation{Department of Physics and Center for Mathematics and Theoretical Physics, National Central University, Chungli, Taiwan 320, Taiwan}%
\affiliation{Institute of Physics, Academia Sinica, Taipei, Taiwan 115, Taiwan}

\date{\today}
\begin{abstract}
Doubly charged Higgs bosons, $H^{\pm\pm}$, are being searched for in the Tevatron experiments.  The most recent search by the D0 collaboration seeks three muons ($\mu^\pm\mu^\pm\mu^\mp$), which are assumed to originate from the pair-production process, $q\overline q\to H^{++}H^{--}$, followed by the decay $H^{\pm\pm} \to \mu^\pm \mu^\pm$.  In this three-lepton ($3{\ell}$) channel there are six distinct signatures for ${\ell}=e$ or $ \mu$.  In the context of the Higgs Triplet Model, we quantify the dependence of the event numbers for the $3{\ell}$ channels on the parameters of the neutrino mass matrix.  It is also shown that the inclusion of the production mechanism $q \overline{q'} \to H^{\pm\pm}H^{\mp}$, followed by the decay $H^\pm\to {\ell}^\pm\nu$, would significantly increase the discovery potential in these channels.
We then provide perspectives on the production of these channels at the Tevatron and LHC.
\end{abstract}
\pacs{14.60.Pq, 14.80.Cp}
\maketitle


\section{Introduction} 
In the last decade, experiments have firmly established that neutrinos have small but finite masses below the eV scale \cite{Fukuda:1998mi}.  Moreover, it is a puzzle why their masses are so much smaller than the charged fermions. These facts motivate physics beyond the Standard Model (SM), which can potentially manifest itself at the CERN Large Hadron Collider (LHC) and/or in low-energy experiments searching for lepton-flavour-violating (LFV) processes \cite{Kuno:1999jp}.  Consequently, models of neutrino mass generation which can be probed at present and forthcoming experiments are of great phenomenological interest.

Among the various viable models, most need to introduce additional heavy neutrinos and/or some extremely high scale in order to naturally explain why the observed neutrinos are so much lighter than the charged fermions, a celebrated example being the Type-I seesaw mechanism with heavy neutrinos of the order of the grand unification scale \cite{Minkowski:1977sc}.  Alternatively, neutrinos could be massless at the tree level, but acquire a mass by radiative corrections \cite{Zee:1980ai}.  However, neutrinos may obtain a mass at tree level from a neutral Higgs boson that acquires a vacuum expectation value (VEV) \cite{Konetschny:1977bn,Schechter:1980gr,Mohapatra:1980yp}.  A particularly simple implementation of this mechanism of neutrino mass generation is the ``Higgs Triplet Model'' (HTM) in which the SM Lagrangian is augmented solely by an $SU(2)$ triplet of scalar particles with hypercharge $Y=2$ \cite{Konetschny:1977bn,Schechter:1980gr}.  Aside from neutrino phenomenology, a distinctive signal of the HTM would be the observation of doubly charged Higgs bosons ($H^{\pm\pm}$), whose mass ($M_{H^{\pm\pm}}$) may be of the order of the electroweak scale.  Such particles can be produced with sizeable rates at hadron colliders in the processes $q\overline q\to H^{++}H^{--}$ \cite{Gunion:1989in, Han:2007bk} and $q\overline {q'}\to H^{\pm\pm}H^{\mp}$ \cite{Dion:1998pw,Akeroyd:2005gt}.  Direct searches have been carried out at the Fermilab Tevatron in the production channel $q\overline q\to H^{++}H^{--}$ and decay mode $H^{\pm\pm}\to \ell^\pm_i \ell^\pm_j$ (where $\ell_{i,j}=e,\mu,\tau$), with mass limits of the order $M_{H^{\pm\pm}} > 110$ -- $150$ GeV \cite{Acosta:2004uj,Abazov:2004au,:2008iy,Aaltonen:2008ip}, assuming a branching ratio (BR) of $100\%$ in a given decay channel.

In the most recent search by the D0 collaboration \cite{:2008iy}, three muons ($\mu^\pm\mu^\pm\mu^\mp$) were searched for in order to reduce the SM backgrounds to an acceptable level.  In fact, there are six distinct $3{\ell}$ signatures for ${\ell}=e$ or $ \mu$, and in the context of the HTM the event numbers can be calculated as a function of the parameters of the neutrino mass matrix.  We perform a quantitative analysis to see which of the six channels provides the most promising detection prospects for $H^{\pm\pm}$ in the HTM.  We analyze how the pattern varies with the underlying parameters of the model.  We also study the impact of the important production mechanism $q \overline{q'} \to H^{\pm\pm} H^{\mp}$, which is omitted in the current searches, but would also contribute to the $3{\ell}$ signatures provided that the singly charged Higgs decays leptonically via $H^\pm\to {\ell}^\pm\nu$.  The paper is organized as follows. In section II the HTM is briefly introduced, and in section III the current search strategy for $H^{\pm\pm}$ at the Tevatron is reviewed. Numerical results are contained in section IV, with conclusions given in section V.

\section{The Higgs Triplet Model} 

The HTM model \cite{Konetschny:1977bn, Schechter:1980gr} is an extension of the SM in which only the scalar sector is augmented with a Higgs triplet.  The model \cite{Konetschny:1977bn, Schechter:1980gr} has the following $SU(2)\otimes U(1)_Y$ gauge-invariant Yukawa interactions:
\begin{equation}
{\cal L} \ni h_{ij}\psi_{iL}^TCi\sigma_2\Delta\psi_{jL} + \mbox{h.c.} ~,
\label{trip_yuk}
\end{equation}
where the triplet Yukawa couplings $h_{ij} (i,j=e,\mu,\tau)$ are complex and symmetric, $C$ is the Dirac charge conjugation operator, $\sigma_2$ is a Pauli matrix, $\psi_{iL}=(\nu_i, l_i)_L^T$ is a left-handed lepton doublet, and $\Delta$ is a $2\times 2$ representation of the $Y=2$ complex triplet fields ($\delta^{++},\delta^+,\delta^0$):
\begin{equation}
\Delta
=\bma{cc}
\delta^+/\sqrt{2}  & \delta^{++} \\
\delta^0       & -\delta^+/\sqrt{2}
\ema ~.
\end{equation}
Note that the mass eigenstate $H^{\pm\pm}$ is entirely composed of the triplet field ($H^{\pm\pm}\equiv \delta^{\pm\pm}$), while $H^\pm$ is predominantly $\delta^\pm$, with a small component of isospin doublet scalar ($\Phi$).  A non-zero Higgs triplet VEV, $\langle \delta^0\rangle = v_\Delta / \sqrt{2}$, gives rise to the following Majorana mass matrix for neutrinos:
\begin{equation}
m_{ij}=2h_{ij}\langle \delta^0\rangle = \sqrt{2}h_{ij}v_{\Delta} ~.
\label{nu_mass}
\end{equation}
This simple expression of tree-level masses for the observed neutrinos is essentially the main motivation for studying the HTM.  It provides a direct connection between $h_{ij}$ and the neutrino mass matrix, which gives rise to phenomenological predictions for processes which depend on $h_{ij}$ \cite{Ma:2000wp}.

The mass matrix $m_{ij}$ for three Dirac neutrinos is diagonalized by the PMNS (Pontecorvo-Maki-Nakagawa-Sakata) matrix $V_\MNS$~\cite{Pontecorvo:1957qd}, for which the standard parametrization is:
\begin{equation}
V_{\rm PMNS}^{} =
\bmaT
c_{12}c_{13}                   & s_{12}c_{13}                  & s_{13}e^{-i\delta} \\
-s_{12}c_{23}-c_{12}s_{23}s_{13}e^{i\delta}  & c_{12}c_{23}-s_{12}s_{23}s_{13}e^{i\delta}  & s_{23}c_{13} \\
s_{12}s_{23}-c_{12}c_{23}s_{13}e^{i\delta}   & -c_{12}s_{23}-s_{12}c_{23}s_{13}e^{i\delta} & c_{23}c_{13}  
\emaT
~,
\end{equation}
where $s_{ij}\equiv\sin\theta_{ij}$ and $c_{ij}\equiv \cos\theta_{ij}$, and $\delta$ is the Dirac phase.  The ranges are chosen as $0\leq\theta_{ij}\leq\pi/2$ and $0\leq\delta<2\pi$.  For Majorana neutrinos (which is the case in HTM), two additional phases appear, and then the mixing matrix $V$ becomes
\begin{eqnarray}
 V = V_{\rm PMNS} \times
     \text{diag}( 1, e^{i\phi_1 /2}, e^{i\phi_2 /2}),
\end{eqnarray}
where $\phi_1$ and $\phi_2$ are referred to as the Majorana phases~\cite{Schechter:1980gr,Bilenky:1980cx} and $-\pi \le \phi_1,\phi_2 < \pi$.  One has the freedom to work in the basis in which the charged lepton mass matrix is diagonal, and then the neutrino mass matrix is diagonalized by $V$.  Using Eq.~(\ref{nu_mass}) one can write the couplings $h_{ij}$ as follows~\cite{Ma:2000wp, Chun:2003ej}:
\begin{equation}
h_{ij}
= \frac{m_{ij}}{\sqrt{2}v_\Delta}
\equiv \frac{1}{\sqrt{2}v_\Delta}
\left[
 V_{\rm PMNS}
 \text{diag}(m_1,m_2 e^{i\phi_1},m_3 e^{i\phi_2})
 V_{\rm PMNS}^T
\right]_{ij} ~.
\label{hij}
\end{equation}
Here $m_1,m_2$ and $m_3$ are the absolute masses of the three neutrinos.  Neutrino oscillation experiments are sensitive to mass-squared differences, $\Delta m_{21}^2$($\equiv m^2_2-m^2_1)$ and $\Delta m_{31}^2$($\equiv m^2_3-m^2_1$).  Since the sign of $\Delta m_{31}^2$ is undetermined at present, distinct patterns for the neutrino mass hierarchy are possible.  The case with $\Delta m^2_{31} >0$ is referred to as {\it normal hierarchy} (NH) where $m_1 < m_2 < m_3$, and the case with $\Delta m^2_{31} <0$ is known as {\it inverted hierarchy} (IH) where $m_3 < m_1 < m_2$.  Information on the mass $m_0$ of the lightest neutrino (either $m_1$ or $m_3$) and the Majorana phases cannot be obtained from neutrino oscillation experiments.  This is because the oscillation probabilities are independent of these parameters, not only in vacuum but also in matter.

In this work we consider $v_\Delta< 0.1$ MeV, which is realized if $h_{ij}$ is larger than the smallest Yukawa coupling in the SM ({\it i.e.,} the electron Yukawa coupling). 
In this case, the leptonic branching ratios (BR's) of $H^{\pm\pm}$ and $H^{\pm}$ are dominant ({\it e.g.}, see \cite{Perez:2008ha}), while $H^{\pm\pm}\to W^\pm W^\pm$ is negligible.\footnote{In the HTM the mass splitting between $M_{H^\pm}$ and $M_{H^{\pm\pm}}$ is caused by a term in the Higgs potential $\lambda_5 \Phi^\dagger \Phi \Delta^\dagger \Delta$, and for small values of the coupling $\lambda_5$ the potentially important decay mode $H^{\pm\pm}\to H^\pm W^*$ \cite{Chakrabarti:1998qy,Akeroyd:2005gt} is very suppressed.}  The BR of $H^{\pm\pm}\to \ell^\pm \ell^\pm$ depends on the six parameters of the neutrino mixing matrix, $V$, (with the dominant uncertainty arising from the unknown Majorana phases, $\phi_1$ and $\phi_2$), the unknown mass of the lightest neutrino ($m_0$), the mass splittings of the neutrinos, and the ignorance of the neutrino mass hierarchy (normal or inverted) \cite{Chun:2003ej}.  Detailed studies of BR$(H^{\pm\pm}\to \ell^\pm \ell^\pm)$ have been performed in \cite{Garayoa:2007fw,Perez:2008ha}.  Notably, BR$(H^\pm\to \ell^\pm \nu)$ (in which the three flavours of neutrinos are summed over) does not depend on the Majorana phases, and the dominant uncertainty is from $m_0$ and the neutrino mass hierarchy \cite{Perez:2008ha}.  Importantly, BR$(H^{\pm\pm}\to \ell^\pm \ell^\pm)\sim 100\%$ and BR$(H^\pm\to \ell^\pm \nu)\sim 100\%$ for a given lepton flavour is not possible in the HTM.  The charged scalars also induce LFV decays such as $\mu\to e\gamma$, $\mu\to eee$ and $\tau\to lll$ \cite{Cuypers:1996ia}, whose rates depend on the parameters of the neutrino mass matrix and the absolute values of $h_{ij}$ \cite{Chun:2003ej, Kakizaki:2003jk, Akeroyd:2009nu}.

\section{Searches for doubly charged Higgs bosons}

Direct searches for $H^{\pm\pm}$ have been carried out at LEP \cite{Abbiendi:2001cr}, the Fermilab Tevatron \cite{Acosta:2004uj,Abazov:2004au,:2008iy,Aaltonen:2008ip} and HERA \cite{Aktas:2006nu}.  The searches in \cite{Abbiendi:2001cr} utilize the production mechanism $e^+e^-\to H^{++}H^{--}$, and the searches in \cite{Acosta:2004uj,Abazov:2004au,:2008iy,Aaltonen:2008ip} assume production via $q\overline q \to H^{++}H^{--}$. Both of these production mechanisms depend on only one unknown parameter, $M_{H^{\pm\pm}}$, while the search strategy in \cite{Aktas:2006nu} depends on both $M_{H^{\pm\pm}}$ and $h_{ij}$.  Production mechanisms which depend on the triplet VEV ($q\overline {q'}\to W^{\pm *}\to W^\mp H^{\pm\pm}$ and fusion via $W^{\pm *} W^{\pm *} \to H^{\pm\pm}$ \cite{Huitu:1996su}) are not competitive with $q\overline q \to H^{++}H^{--}$ at the energies of the Tevatron, but can be the dominant source of $H^{\pm\pm}$ at the LHC if $v_{\Delta}={\cal O}$ (1 GeV) and $M_{H^{\pm\pm}}> 500$ GeV.  All searches assume the leptonic decay mode $H^{\pm\pm}\to \ell^\pm \ell^\pm$, for which there are six possibilities ($ee,\mu\mu,\tau\tau,e\mu,e\tau,\mu\tau$).  For the case of BR$(H^{\pm\pm}\to \ell^\pm \ell^\pm)<100\%$, the expected number of $H^{\pm\pm}\to \ell^\pm \ell^\pm$ events scales linearly in BR (for searches for a single pair of same-sign leptons) or quadratically in BR (for searches which require a third lepton or more).  Explicit expressions will be given below.  Thus, the mass limits for $M_{H^{\pm\pm}}$ can be considerably weakened for the (more realistic) scenarios in which BR$(H^{\pm\pm}\to \ell^\pm \ell^\pm)<100\%$. The search strategy at LEP \cite{Abbiendi:2001cr} required four leptons.


The CDF collaboration searched for three final states, $H^{\pm\pm}\to e^\pm e^\pm, e^\pm \mu^\pm, \mu^\pm\mu^\pm$, requiring at least one pair of same-sign leptons with high invariant mass \cite{Acosta:2004uj}.  The integrated luminosity used was $0.24$ fb$^{-1}$ and the mass limits $M_{H^{\pm\pm}}>133,113,136$ GeV were obtained for the decay channels $H^{\pm\pm}\to e^\pm e^\pm,e^\pm \mu^\pm, \mu^\pm\mu^\pm$, respectively, assuming BR=$100\%$ in a given channel.  The D0 collaboration \cite{Abazov:2004au,:2008iy} searched for $H^{\pm\pm}\to \mu^\pm\mu^\pm$.  We note here that the main difference between these searches by D0 is the requirement \cite{:2008iy} of a third $\mu$ of opposite sign to the two same-sign $\mu$, the latter assumed to originate from the decay of one of the pair-produced $H^{\pm\pm}$.  This extra requirement suppresses backgrounds from $\gamma/Z\to \mu^+\mu^-$ and multijets, which were less than one event for the integrated luminosity of $0.11$ fb$^{-1}$ used in \cite{Abazov:2004au}, but became non-negligible for the search in \cite{:2008iy} with $1.1$ fb$^{-1}$.  The requirement of a third lepton is necessary for the future Tevatron searches in order to reduce the SM backgrounds.  But such a cut means that signal events would be lost for the realistic case of BR$(H^{\pm\pm}\to \ell^\pm \ell^\pm)< 100\%$.

All the above searches at the Tevatron assume {\it only} the production mechanism $q\overline q\to \gamma^*,Z^*\to H^{++}H^{--}$.  However, the process $q\overline {q'}\to W^*\to H^{\pm\pm}H^\mp$ \cite{Dion:1998pw,Akeroyd:2005gt} has a cross section comparable to that of $q\overline q\to H^{++}H^{--}$ for $M_{H^\pm}\sim M_{H^{\pm\pm}}$ at hadron colliders, and thus the former will also contribute to the search for $H^{\pm\pm}$.  In Ref.~\cite{Akeroyd:2005gt}, it is suggested that the search potential at hadron colliders can be improved by considering the following inclusive single $H^{\pm\pm}$ cross section ($\sigma_{H^{\pm\pm}}$):
\begin{eqnarray}
\sigma_{H^{\pm\pm}} &=& \sigma(q\overline q\to \gamma^*,Z^*\to H^{++}H^{--}) 
\nonumber \\
&& + 2\sigma(q\overline {q'}\to W^*\to H^{++}H^-) ~,
\label{single_prod}
\end{eqnarray}
where the factor of two arises from the charge conjugate process $q'\overline q \to W^*\to H^{--}H^+$.  The cross section for at least one pair of same-sign leptons $\ell^\pm \ell^\pm$ (we focus on $\ell=e$ or $\mu$, for which hadron colliders have the greatest discovery potential) is given by:
\begin{eqnarray}
\sigma_{\ell\ell} &=& 
\sigma(p\overline p\to H^{++}H^{--}) \times
{\cal B}_{\ell\ell} (2 - {\cal B}_{\ell\ell}) \nonumber \\
&&
+ 2\sigma(p\overline p\to H^{++}H^{-})\times {\cal B}_{\ell\ell} ~,
\label{cross-section}
\end{eqnarray}
where ${\cal B}_{\ell\ell} \equiv$ BR$(H^{\pm\pm}\to \ell^\pm \ell^\pm)$.  For illustration, we take $M_{H^{\pm\pm}}= 150$ GeV and $M_{H^{\pm\pm}}=M_{H^\pm}$.  Then one has $2\sigma (p \overline p \to H^{++}H^{-}) \simeq 1.2\sigma(p\overline p\to H^{++}H^{--})$ at the energy of the Tevatron \cite{Akeroyd:2005gt}, and Eq.~(\ref{cross-section}) simplies to:
\begin{equation}
\sigma_{\ell\ell} = \sigma(p\overline p\to H^{++}H^{--}) \times
{\cal B}_{\ell\ell} (3.2 - {\cal B}_{\ell\ell}) ~.
\label{two-lepton_cross}
\end{equation}
In the following, we normalize the production rate by $\sigma(p\overline p\to H^{++}H^{--})$ and denote the corresponding quantity with a hat.  Therefore,
\begin{equation}
\hat\sigma_{\ell\ell} = {\cal B}_{\ell\ell} (3.2 - {\cal B}_{\ell\ell}) ~.
\label{two-lepton}
\end{equation}
The factor $3.2$ in the above expression is replaced by $3.8$ for the energy of the LHC and taking $M_{H^{\pm\pm}}=M_{H^\pm}=250$ GeV.  The dominant term in this expression is linear in ${\cal B}_{\ell\ell}$, and there is no dependence on BR$(H^\pm\to \ell^\pm \nu)$.  The first searches for $H^{\pm\pm}$ at the Tevatron \cite{Acosta:2004uj,Abazov:2004au} looked for at least two same-sign leptons, and such a strategy probed the cross section in Eq.~(\ref{two-lepton}).  Mass limits on $M_{H^{\pm\pm}}$ were obtained, neglecting the contribution from $q\overline {q'}\to W^*\to H^{\pm\pm}H^\mp$.  However, with the larger data samples now available, the requirement of a pair of same-sign leptons is not sufficient to reduce the SM backgrounds to a negligible level, which is necessary in order to probe larger values of $M_{H^{\pm\pm}}$.  The requirement of a third lepton is needed in order to improve the limits on $M_{H^{\pm\pm}}$ which were obtained in \cite{Acosta:2004uj,Abazov:2004au}, and hence the most recent Tevatron search \cite{:2008iy} requires a third $\mu$ of opposite sign to a pair of same-sign $\mu$.

On requiring a third lepton there are six distinct signatures for $\ell=e$ or $\mu$: $e^\pm e^\pm e^\mp$, $e^\pm e^\pm \mu^\mp$, $e^\pm \mu^\pm e^\mp$, $e^\pm \mu^\pm\mu^\mp$, $\mu^\pm \mu^\pm e^\mp$ and $\mu^\pm \mu^\pm \mu^\mp$.  The explicit expressions for the cross sections of (at least) three leptons ($\ell^\pm \ell^\pm \ell^\mp$) are given below, where the first two leptons in the subscript of $\hat\sigma_{\ell \ell \ell}$ have the same sign and the third lepton is of opposite sign:
\begin{eqnarray}
\hat\sigma_{eee} &=& 
{\cal B}_{ee} \left[
{\cal B}_{ee} + 2({\cal B}_{e\mu}+{\cal B}_{e\tau})+1.2{\cal B}_{e\nu} \right] ~,
\label{three-lepton-start} \\
\hat\sigma_{ee\mu} &=& 
{\cal B}_{ee} \left[
2({\cal B}_{\mu\mu} + {\cal B}_{e\mu}+{\cal B}_{\mu\tau})+1.2{\cal B}_{\mu\nu} \right]
~, \\
\hat\sigma_{e\mu e} &=& 
{\cal B}_{e\mu} \left[
{\cal B}_{e\mu} + 2({\cal B}_{ee} + {\cal B}_{e\tau}) + 1.2{\cal B}_{e\nu} \right] 
~, \\
\hat\sigma_{e\mu \mu} &=& 
{\cal B}_{e\mu}  \left[ {\cal B}_{e\mu} + 2({\cal B}_{\mu\mu} + {\cal B}_{\mu\tau}) + 1.2{\cal B}_{\mu\nu} \right] ~, \\
\hat\sigma_{\mu \mu e} &=&
{\cal B}_{\mu\mu} \left[
2({\cal B}_{ee} + {\cal B}_{e\mu} + {\cal B}_{e\tau}) + 1.2{\cal B}_{e\nu} \right] 
~, \\
\hat\sigma_{\mu\mu\mu} &=&
{\cal B}_{\mu\mu} \left[ {\cal B}_{\mu\mu} + 2({\cal B}_{e\mu} + {\cal B}_{\mu\tau}) + 1.2{\cal B}_{\mu\nu} \right] ~.
\label{three-lepton-end}
\end{eqnarray}
Here ${\cal B}_{\ell\nu}\equiv$ BR$(H^\pm\to \ell^\pm \nu)$ and the factors of two are combinatorial.  The coefficient $1.2$ in front of ${\cal B}_{\ell\nu}$ is replaced by $\sim 1.8$ at the LHC and taking $M_{H^{\pm\pm}}=M_{H^\pm}=250$ GeV.  The magnitude of this coefficient has a small dependence on $M_{H^{\pm\pm}}$, and for the LHC it increases to $\sim 2.0$ for $M_{H^{\pm\pm}}=1000$ GeV \cite{Akeroyd:2005gt}.  In the above expressions, we do not include contributions to $\hat\sigma_{\ell\ell\ell}$ originating from combinations such as ${\cal B}_{\ell \tau} {\cal B}_{\ell \tau}$, in which one $\tau$ decays leptonically via $\tau\to \ell\nu\nu$.  The magnitude of such contributions are suppressed by the branching ratio of $\tau\to \ell\nu\nu \sim 17\%$ for $\ell=e$ and $\mu$.  Moreover, any $\ell$ from $\tau\to \ell\nu\nu$ would be less energetic than an $\ell$ arising directly from the decay of $H^{\pm\pm}$, and thus it would be less likely to pass the cut on the tranverse momentum of $\ell$. Given these suppression factors, we neglect contributions of the type ${\cal B}_{\ell \tau} {\cal B}_{\ell \tau}$, which would slightly increase the magnitude of $\hat\sigma_{\ell\ell\ell}$.

In Ref.~\cite{delAguila:2008cj}, a simulation was performed for an inclusive three-lepton signature at the LHC, in which $e$ and $\mu$ were not distinguished. Such an inclusive channel has the advantage of maximizing the sensitivity to $M_{H^{\pm\pm}}$ for a given integrated luminosity.  The three-lepton signature in \cite{delAguila:2008cj} requires {\sl exactly} three leptons, and so it differs from that defined in Eqs.~(\ref{three-lepton-start}) to (\ref{three-lepton-end}).  Careful attention was given to the contribution of $\tau\to \ell\nu\nu$, where ${\cal B}_{\ell \tau} {\cal B}_{\ell \tau}$ with one tau decaying leptonically and one $\tau$ decaying hadronically also give the necessary three leptons.  It was concluded that this inclusive three-lepton signature offered greater discovery potential for $H^{\pm \pm}$ in the HTM than the signatures of exactly two or four leptons.  This is mainly because the three-lepton signature has the extra contribution from $pp\to H^{\pm\pm}H^\mp$ (which is not relevant for the four lepton signature), and has a SM background similar in magnitude to that for the four-lepton signature in the region of high invariant mass of $\ell\ell$.

Ref.~\cite{delAguila:2008cj} also acknowledges that the magnitude of $\hat\sigma_{\ell\ell\ell}$ for the six exclusive channels depends on the neutrino parameters in the HTM, and such channels should be investigated separately if there is any signal in the inclusive channel.  In the next section, we will investigate the magnitude of $\hat\sigma_{\ell\ell\ell}$ as defined by Eqs.(\ref{three-lepton-start}) to (\ref{three-lepton-end}).  At present, the Tevatron has only searched in one exclusive three-lepton channel, $\mu^\pm\mu^\pm\mu^\mp$ \cite{:2008iy}.  The SM backgrounds to the six exclusive channels (at either the Tevatron or LHC) will not be identical, although one expects the backgrounds to be very similar for the region of large dilepton invariant mass (say $M_{H^{\pm\pm}}> 250$ GeV, which cannot be probed at the Tevatron but is relevant for the LHC), as discussed in \cite{delAguila:2008cj}. Morover, efficiencies for lepton tagging will not be the same for $e^\pm$ and $\mu^\pm$.  As an example, the search by the CDF collaboration for $ee$, $e\mu$ and $\mu\mu$ in the two-lepton (or more) channel \cite{Acosta:2004uj} (defined by Eq.~(\ref{two-lepton})) showed similar sensitivity to the decay modes $H^{\pm\pm}\to e^\pm e^\pm$ and $H^{\pm\pm}\to \mu^\pm \mu^\pm$ (with mass limits $M_{H^{\pm\pm}}> 133$ GeV and $M_{H^{\pm\pm}}> 136$ GeV respectively), but slightly inferior sensitivity to $H^{\pm\pm}\to e^\pm \mu^\pm$ (mass limit $M_{H^{\pm\pm}}> 113$ GeV).

In Ref.~\cite{Perez:2008ha}, a simulation at the LHC was performed for the production channel $pp\to H^{\pm\pm}H^\mp$ followed by the decays $H^{\pm\pm}\to \ell^\pm \ell^\pm$ and $H^\pm\to \ell^\pm\nu$, where both $e$ and $\mu$ contributions are summed together in an inclusive approach like that in \cite{delAguila:2008cj}.  The difference between \cite{Perez:2008ha} and \cite{delAguila:2008cj} is that the former imposes a cut on missing energy (which originates from $H^\pm\to \ell^\pm\nu$) in order to remove the contribution from $pp\to H^{++}H^{--}$ and isolate the contribution from $pp\to H^{\pm\pm}H^\mp$.  Such a strategy is therefore probing the last term in Eqs.~(\ref{three-lepton-start}) to (\ref{three-lepton-end}) and summing over $e^\pm$ and $\mu^\pm$.  This approach probes the vertex $H^{\pm\pm}H^\mp W^\pm$, which is present in the HTM but not in models with $SU(2)$ singlet scalars. Simulations at the LHC for $pp\to H^{++}H^{--}$ alone have been carried out in \cite{Han:2007bk, Perez:2008ha, Hektor:2007uu}.

\section{Numerical Analysis}
\begin{figure}[t]
\begin{center}
\includegraphics[origin=c, angle=0, scale=0.52]{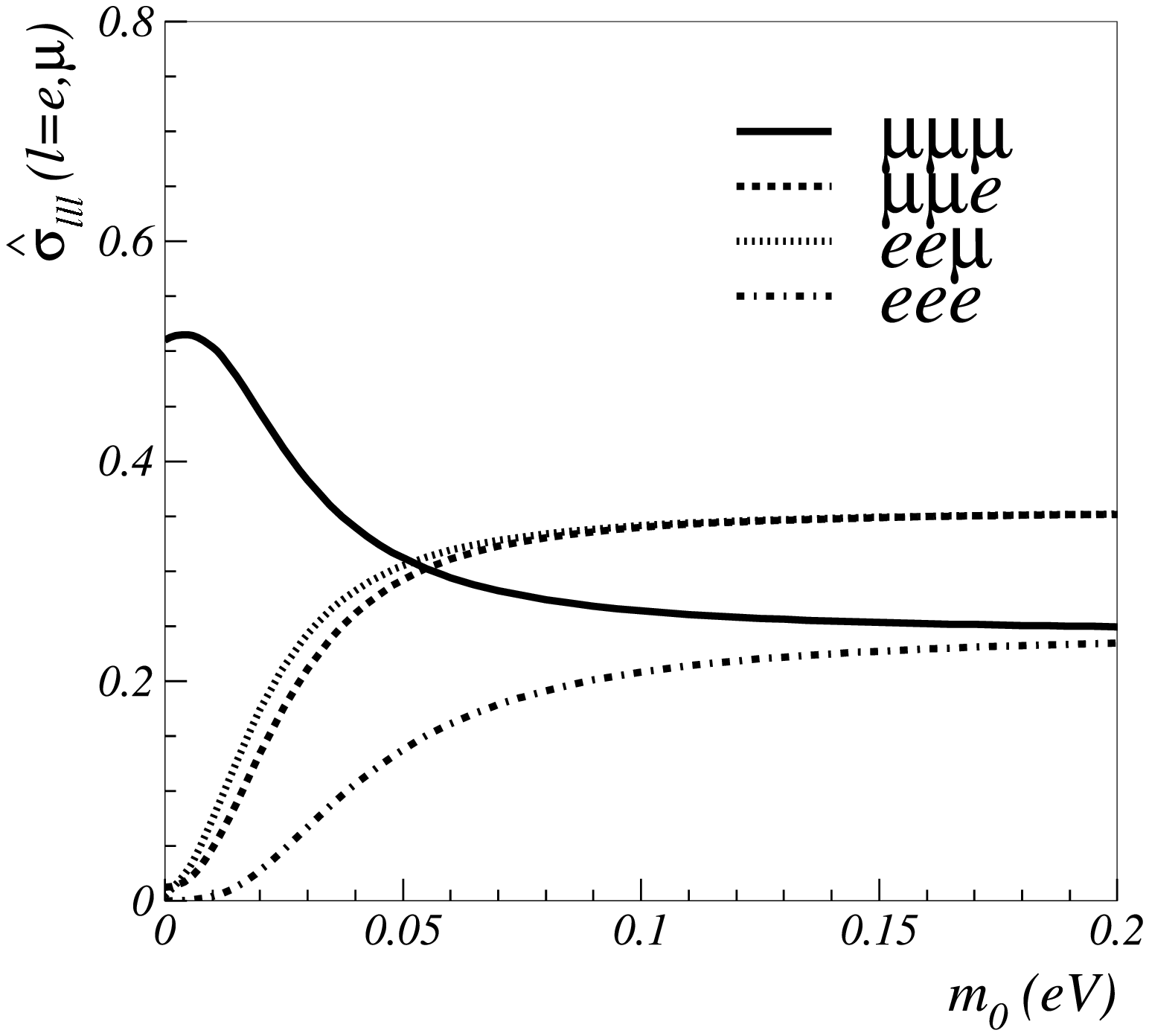}
\includegraphics[origin=c, angle=0, scale=0.52]{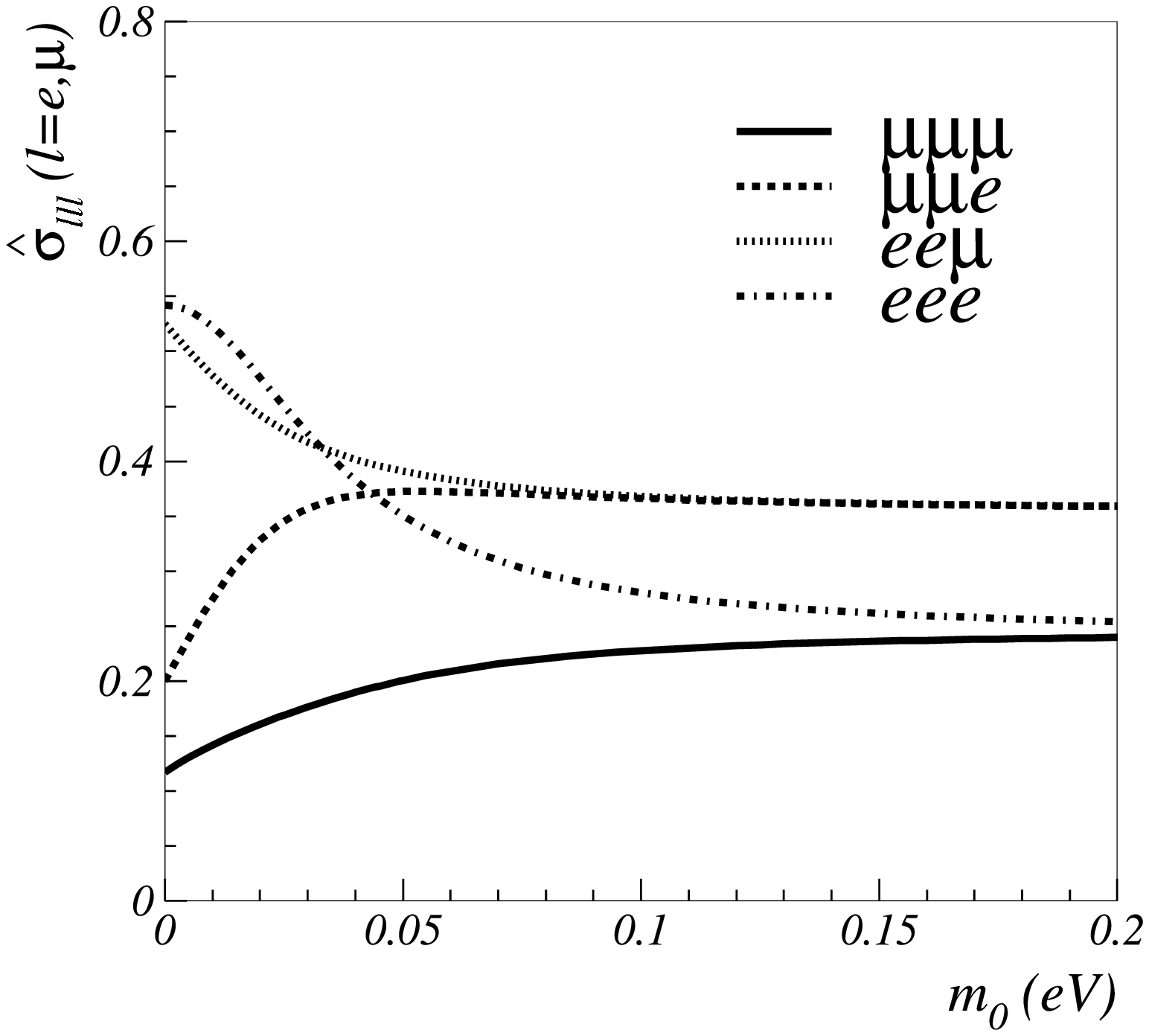}
\caption{The cross section for three leptons, $\hat\sigma_{\ell \ell \ell}$, as a function of $m_0$ for the normal neutrino mass hierarchy (left panel) and the inverted neutrino mass hierarchy (right panel) normalized to $\sigma(pp\to H^{++}H^{--})$ (for the energy of the Tevatron and $M_{H^{\pm\pm}}=150$ GeV).  The curves for $\ell \ell \ell=\mu\mu\mu, \mu\mu e, ee\mu, eee$, where the first two leptons have the same charge ($++$ or $--$), are drawn in solid, dashed, dash-dotted, and dotted lines, respectively.  The lines for $e \mu e$ and $e\mu\mu$ are not shown because $\sigma_{\ell \ell \ell}\sim 0$.  The Majorana phases are set to zero ($\phi_1=\phi_2=0$), and the neutrino oscillation parameters are fixed as: $\Delta m^2_{21} = 8\times 10^{-5} {\rm eV}^2$, $|\Delta m^2_{31}| = 2.5\times 10^{-3} {\rm eV}^2$, $\sin^22\theta_{12}\simeq 0.8$, $\sin^2\theta_{23}=0.5$, and $\sin^22\theta_{13}=0$.}
\label{fig:m0}
\end{center}
\end{figure}

In this section we present our numerical study of the magnitudes of the three-lepton cross sections, $\hat\sigma_{\ell \ell \ell}$, defined by Eqs.~(\ref{three-lepton-start}) to (\ref{three-lepton-end}).  As the primary uncertainties come from the lightest neutrino mass $m_0$ and the Majorana phases $phi_{1,2}$, here we take for definiteness $\Delta m^2_{21} = 8\times 10^{-5} {\rm eV}^2$, $|\Delta m^2_{31}| =\simeq 2.5\times 10^{-3} {\rm eV}^2$, $\sin^22\theta_{12} = 0.8$, $\sin^2\theta_{23}=0.5$, and $\sin^22\theta_{13}=0$.  The Dirac phase $\delta$ is irrelevant here.

\begin{figure}[t]
\begin{center}
\includegraphics[origin=c, angle=0, scale=0.52]{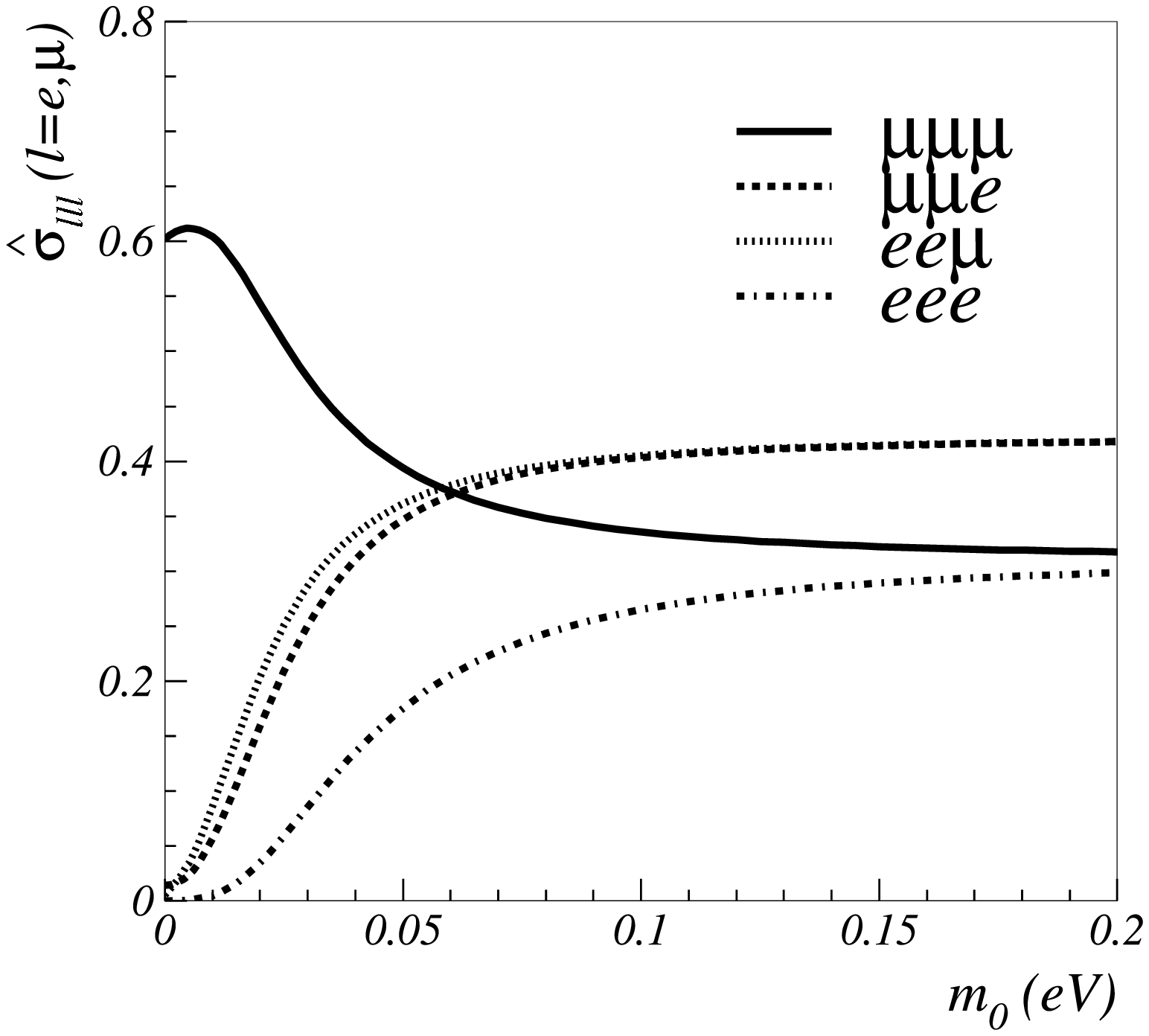}
\includegraphics[origin=c, angle=0, scale=0.52]{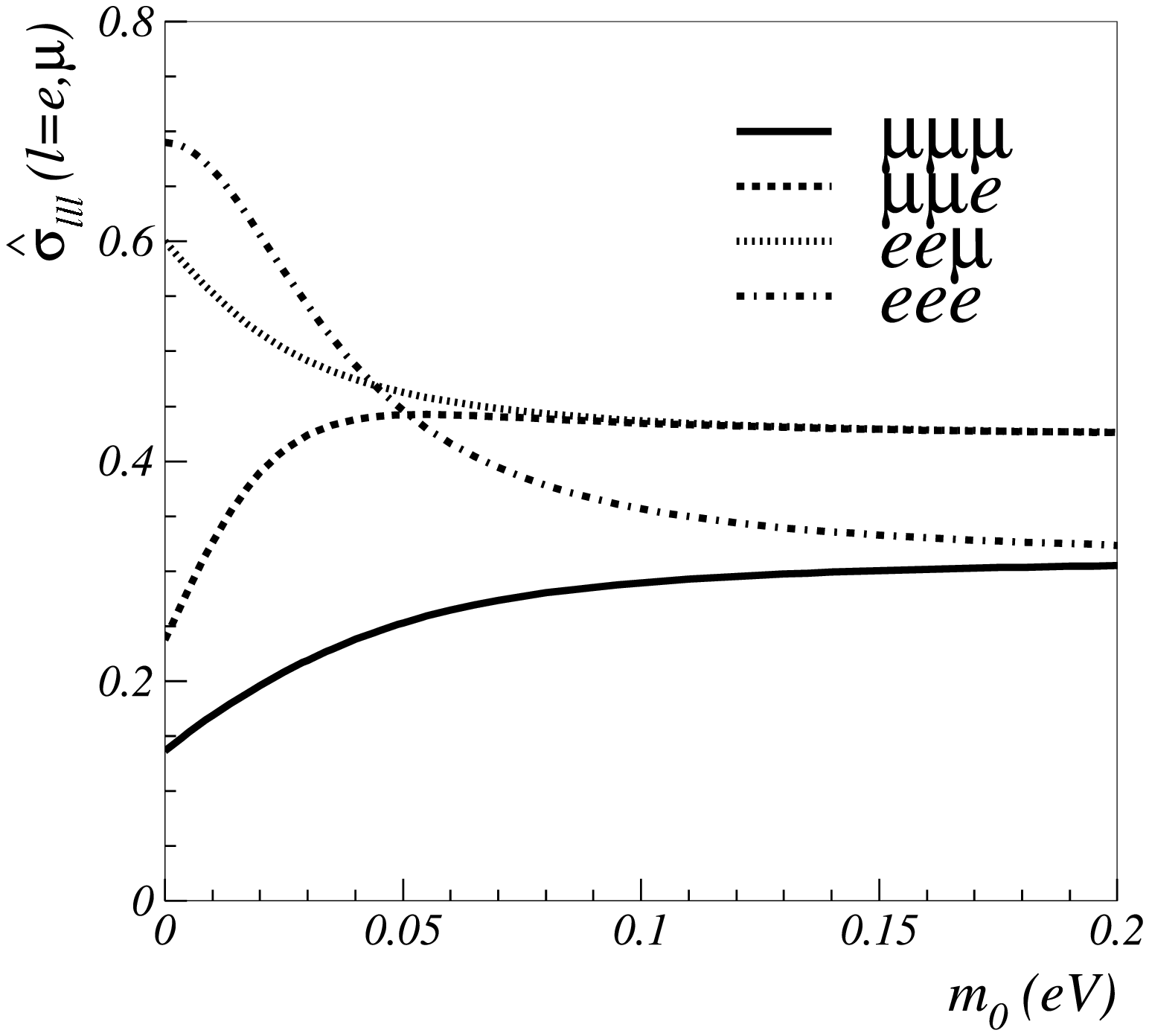}
\caption{Same as Fig.~\ref{fig:m0} but for the LHC, and taking the coefficient of ${\cal B}_{\ell\nu}$ in Eqs.~(\ref{three-lepton-start}) to (\ref{three-lepton-end}) to be 1.8 (correponding to $M_{H^{\pm\pm}}=M_{H^\pm}=250$ GeV).}
\label{fig:lhcm0}
\end{center}
\end{figure}

\begin{figure}[t]
\begin{center}
\includegraphics[origin=c, angle=0, scale=0.52]{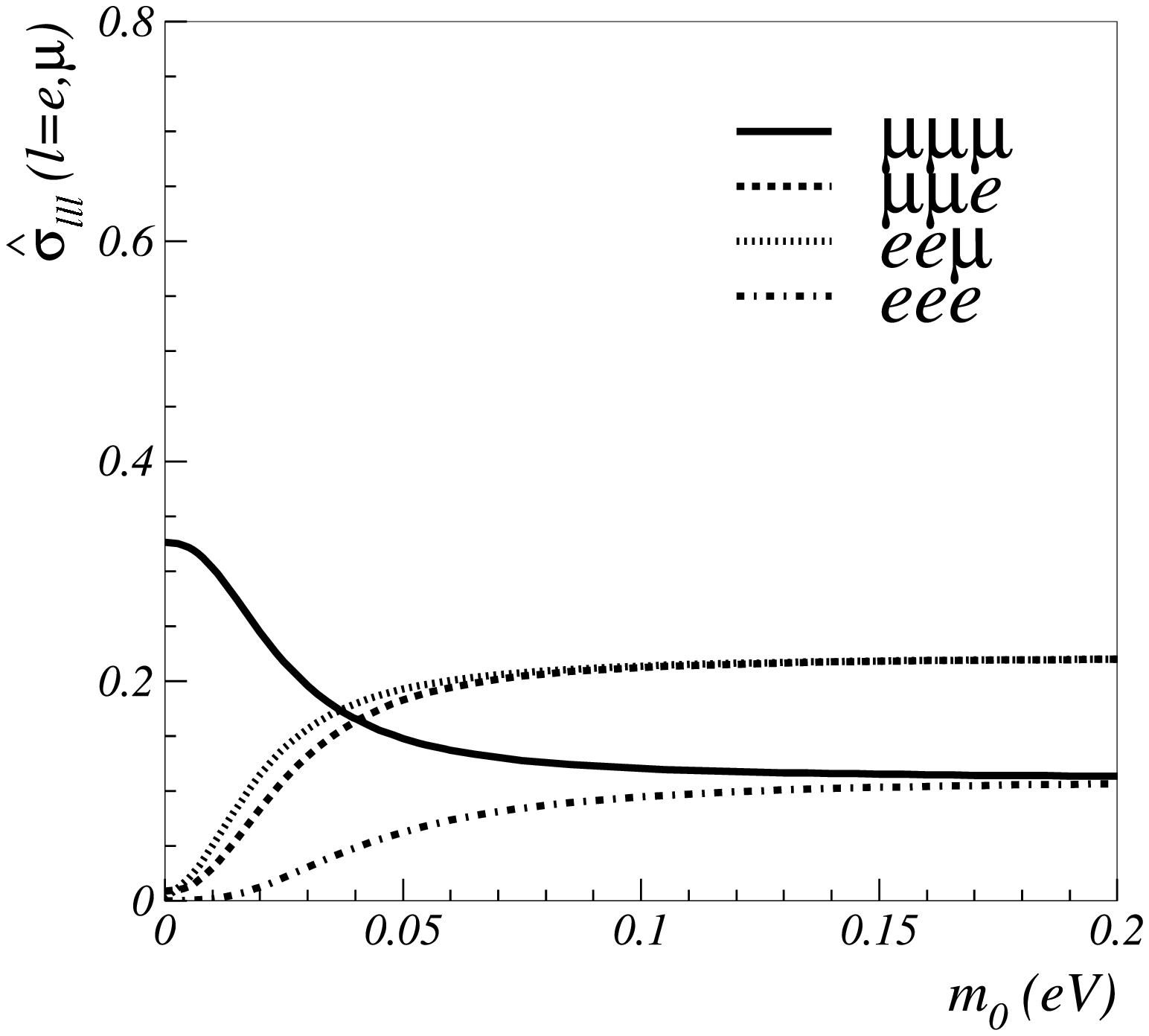}
\includegraphics[origin=c, angle=0, scale=0.52]{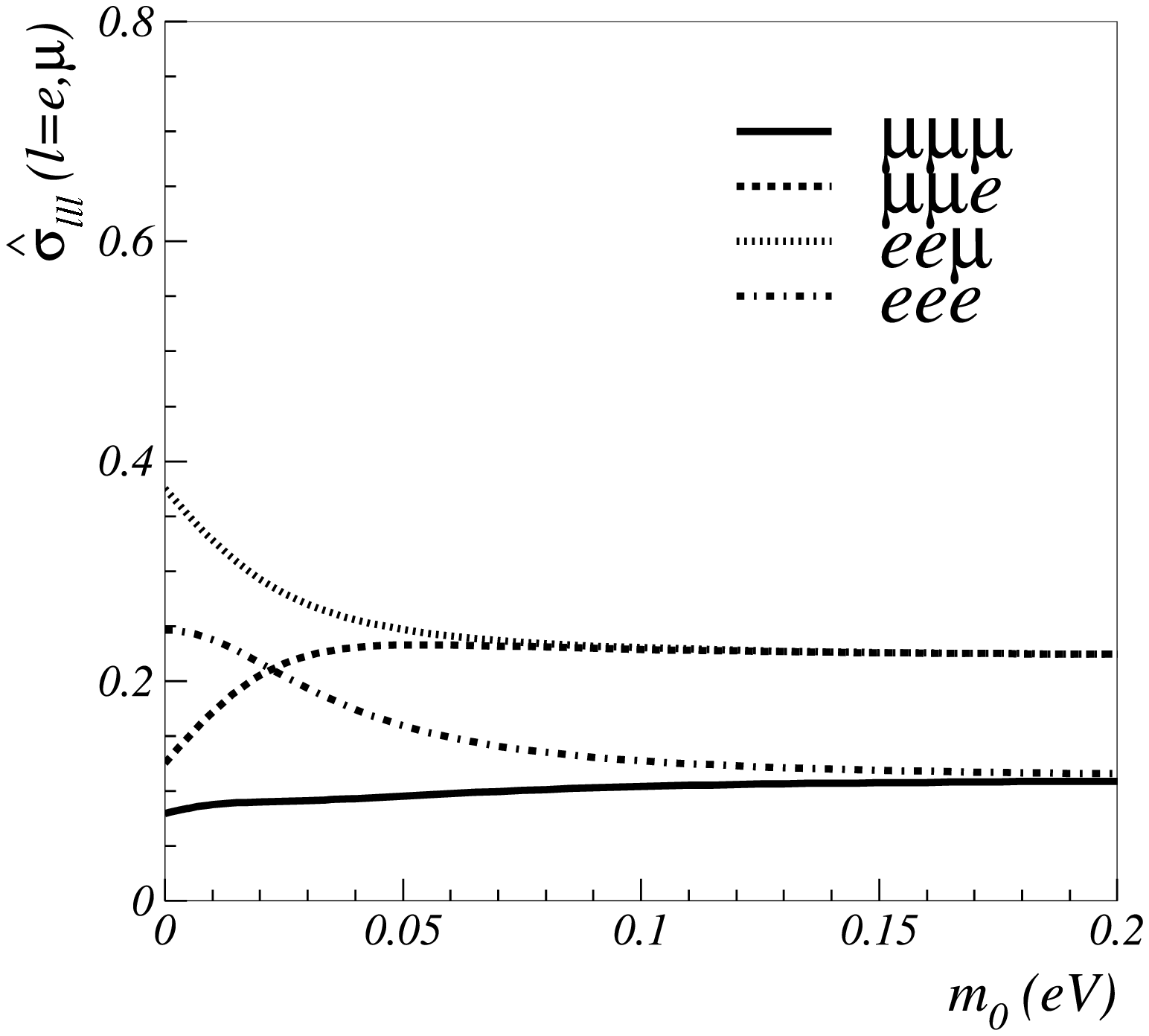}
\caption{Same as Fig.~\ref{fig:m0}, but taking the coefficient of ${\cal B}_{\ell\nu}$ in Eqs.~(\ref{three-lepton-start}) to (\ref{three-lepton-end}) to be 0 ({\it i.e.}, neglecting the contribution from $pp\to H^{\pm\pm}H^{\mp}$).  This figure applies to both the Tevatron and LHC.}
\label{fig:no_sing_prod}
\end{center}
\end{figure}

In Fig.~\ref{fig:m0}, $\hat\sigma_{\ell \ell \ell}$ is plotted as a function of $m_0$ for the energy of the Tevatron.  The left (right) panel is for the normal (inverted) neutrino mass hierarchy.  When making these two plots, we assume $M_{H^{\pm\pm}} = M_{H^\pm} = 150$ GeV ({\it i.e.}, the coefficient of the term ${\cal B}_{\ell\nu}$ is 1.2), set the Majorana phases to zero ($\phi_1=\phi_2=0$), and fix the neutrino oscillation parameters as described above.  In Fig.~\ref{fig:lhcm0}, the corresponding results for the LHC are shown, taking $M_{H^{\pm\pm}} = M_{H^\pm} = 250$ GeV ({\it i.e.}, the coefficient of the term ${\cal B}_{\ell\nu}$ is 1.8).  The curves in Fig.~\ref{fig:m0} and Fig.~\ref{fig:lhcm0} have the same qualitative behaviour, with the numerical differences being caused by the coefficient of the term ${\cal B}_{\ell\nu}$.  Note that no channel reaches a value of $\hat\sigma_{\ell \ell \ell}=1$, and the largest value is $\hat\sigma_{\ell \ell \ell} \sim 0.7$.  This is true even when one varies the Majorana phases (see later).  In the case of the normal mass hierarchy, the dominant channel is to three muons ($\hat\sigma_{\mu\mu\mu}$) when $m_0 \lsim 0.05$ eV. However, in the case of the inverted mass hierarchy the dominant channels in this mass range are $eee$ and $ee\mu$. When $m_0 > 0.05$ GeV (corresponding to quasi-degenerate neutrinos), the dominant channels become $\mu\mu e$ and $ee\mu$ and they saturate around $35\%$.  In both hierarchies, $\hat\sigma_{e \mu e}$ and $\hat\sigma_{e\mu\mu}$ are
 numerically tiny and are not shown, the reason being that this choice of parameters gives ${\cal B}_{e\mu}\sim 0$.  In order to show the impact of the contribution from $pp\to H^{\pm\pm}H^{\mp}$, in Fig.~\ref{fig:no_sing_prod} the cross sections $\hat\sigma_{\ell\ell\ell}$ are plotted with the coefficient of the term ${\cal B}_{\ell\nu}$ set to zero.  In this scenario, $\hat\sigma_{\ell\ell\ell}$ is the same at both the Tevatron and LHC.  From Fig.~\ref{fig:no_sing_prod} it is clear that the numerical values of $\hat\sigma_{\ell\ell\ell}$ have dropped significantly with respect to Fig.~\ref{fig:m0} and Fig.~\ref{fig:lhcm0}, and thus the impact of $pp\to H^{\pm\pm}H^{\mp}$ is seen to be important.  Comparing Fig.~\ref{fig:lhcm0} and Fig.~\ref{fig:no_sing_prod}, the large relative enhancement of $\hat\sigma_{eee}$ is also due to the sizeable ${\cal B}_{e\nu}$ ($\sim 50\%$) in the IH scenario for $m_0 < 0.05$ GeV \cite{Perez:2008ha}.

\begin{figure}[t]
\begin{center}
\includegraphics[origin=c, angle=0, scale=0.46]{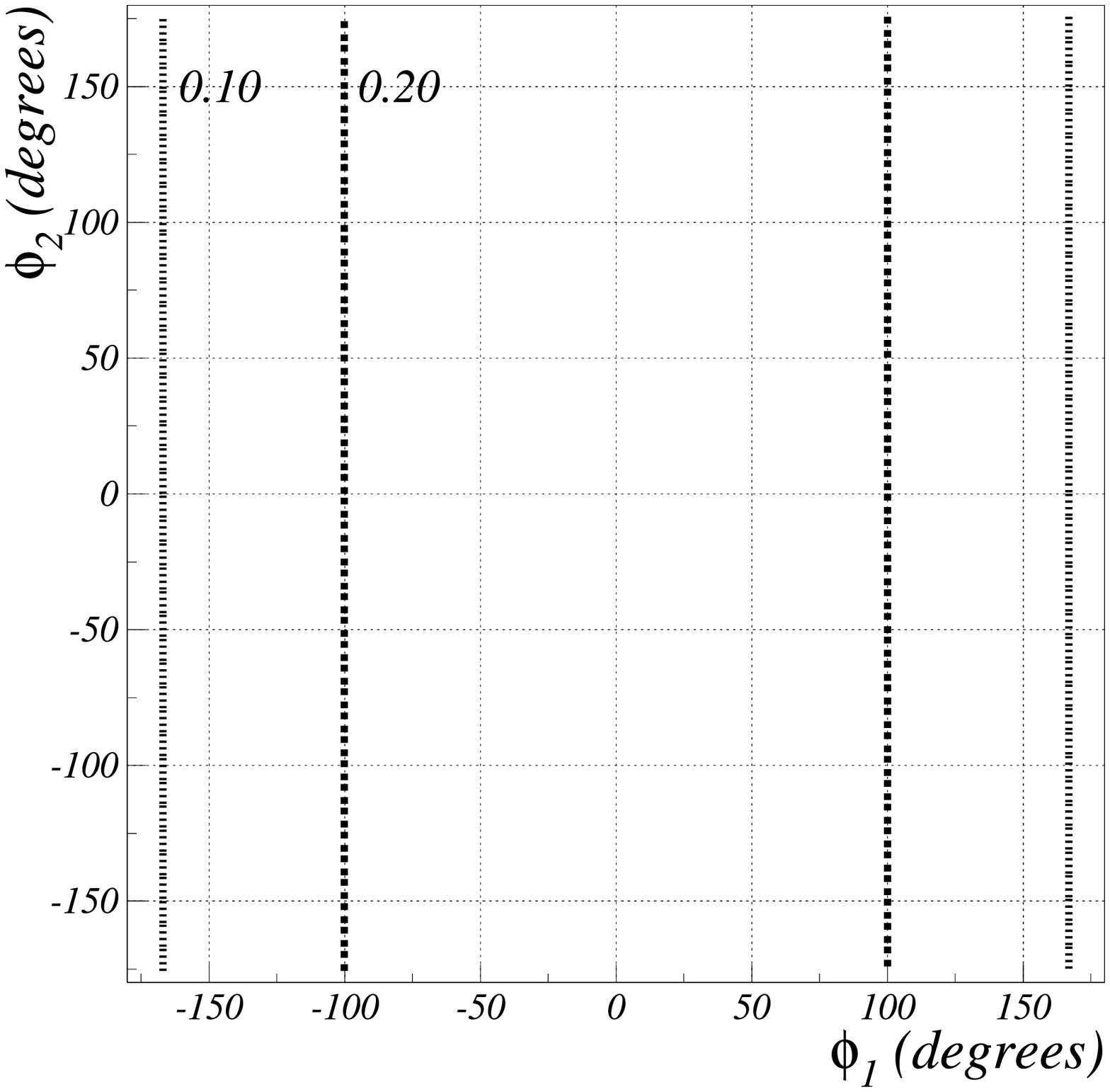}
\includegraphics[origin=c, angle=0, scale=0.46]{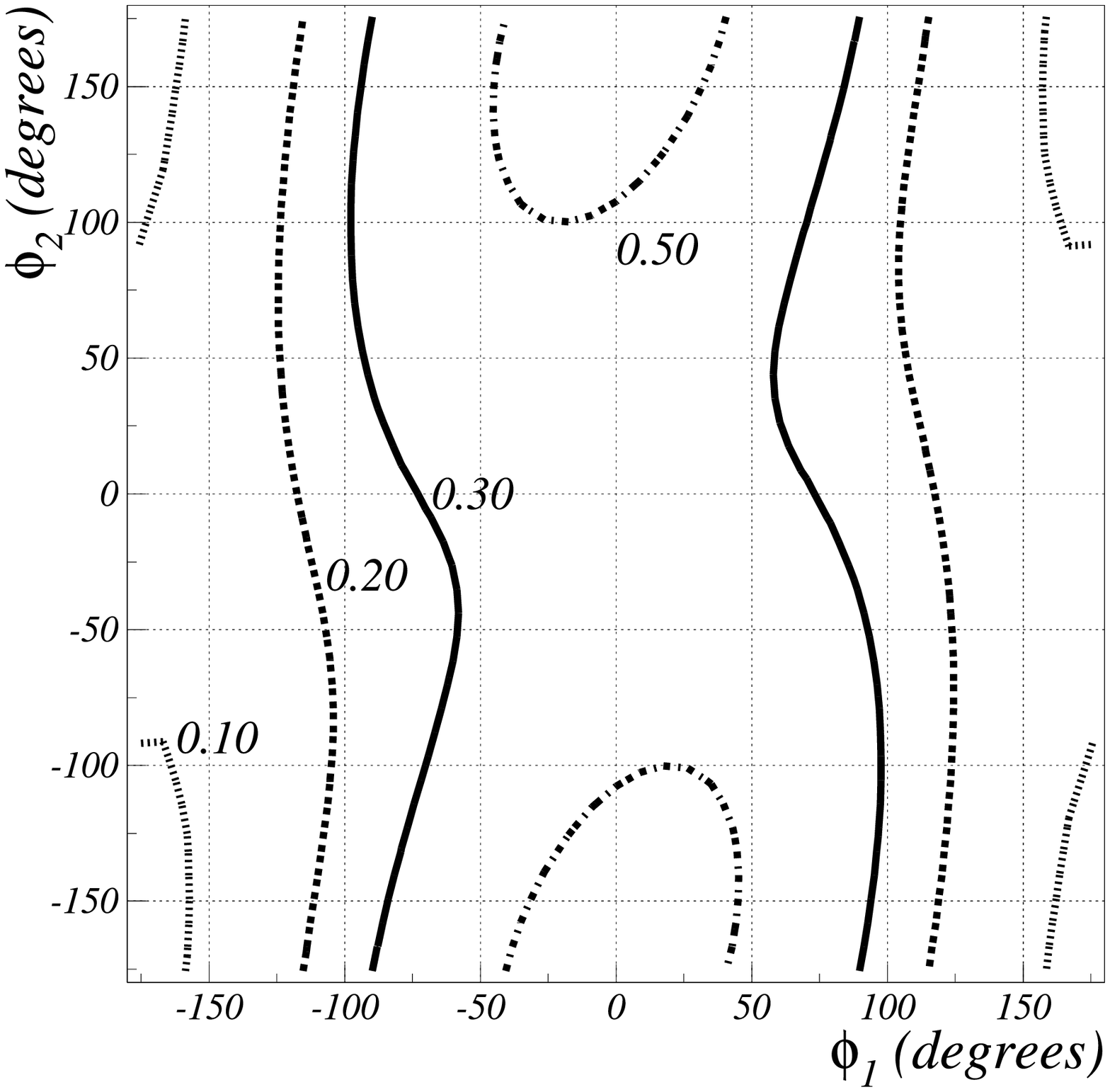}
\caption{Contours of $\hat\sigma_{eee}$ (left), $\hat\sigma_{ee\mu}$ (right) in the $\phi_1$-$\phi_2$ plane for $m_0=0.2$ eV (quasi-degenerate neutrinos).  Other parameters are fixed as in Fig.~\ref{fig:m0}.}
\label{fig:majorana1}
\end{center}
\end{figure}
\begin{figure}[t]
\begin{center}
\includegraphics[origin=c, angle=0, scale=0.46]{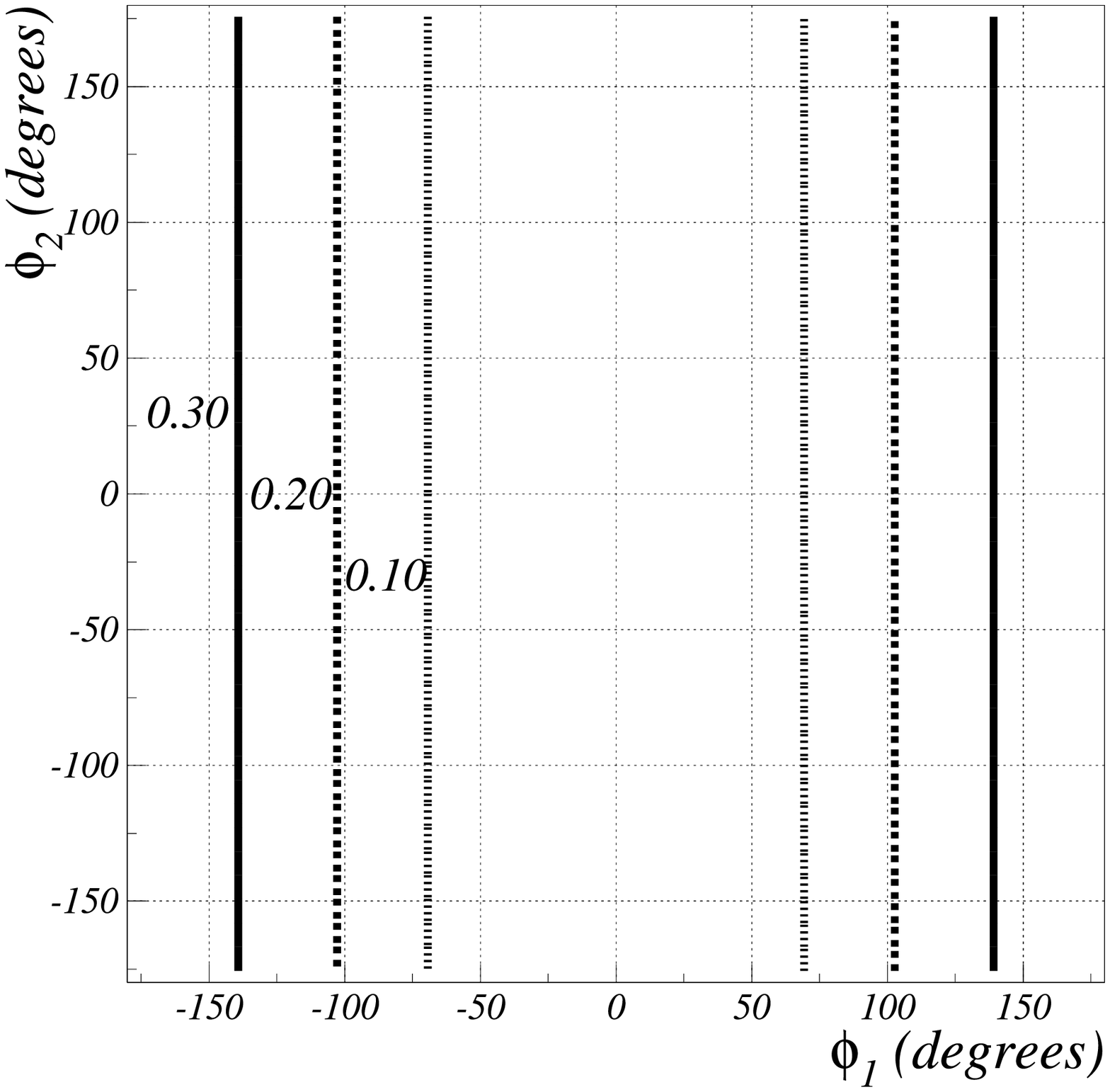}
\includegraphics[origin=c, angle=0, scale=0.46]{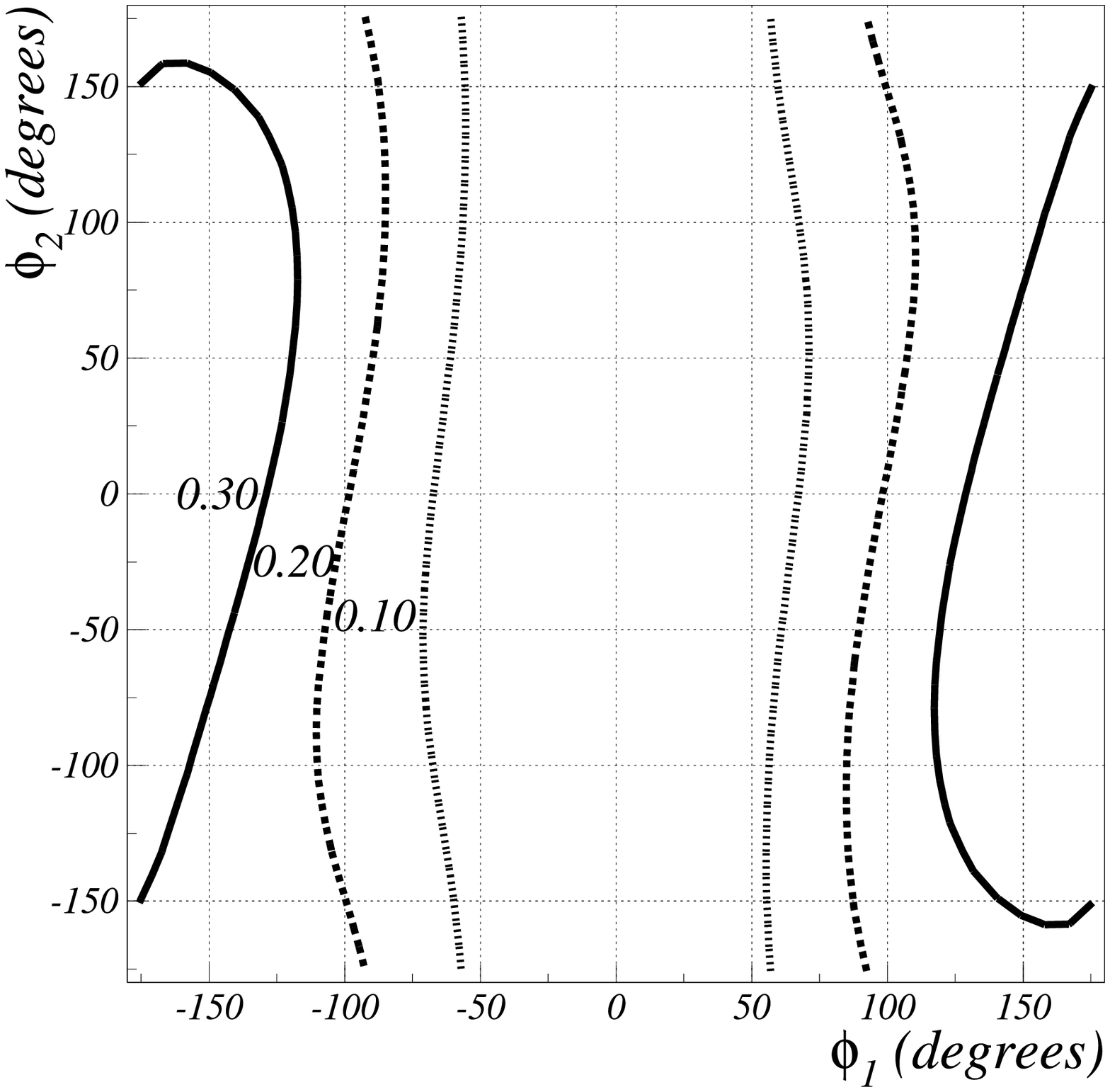}
\caption{Contours of $\hat\sigma_{e\mu e}$ (left), $\hat\sigma_{e\mu \mu}$ (right) in the $\phi_1$-$\phi_2$ plane for $m_0=0.2$ eV (quasi-degenerate neutrinos). Other parameters are fixed as in Fig.~\ref{fig:m0}.}
\label{fig:majorana2}
\end{center}
\end{figure}
\begin{figure}[t]
\begin{center}
\includegraphics[origin=c, angle=0, scale=0.46]{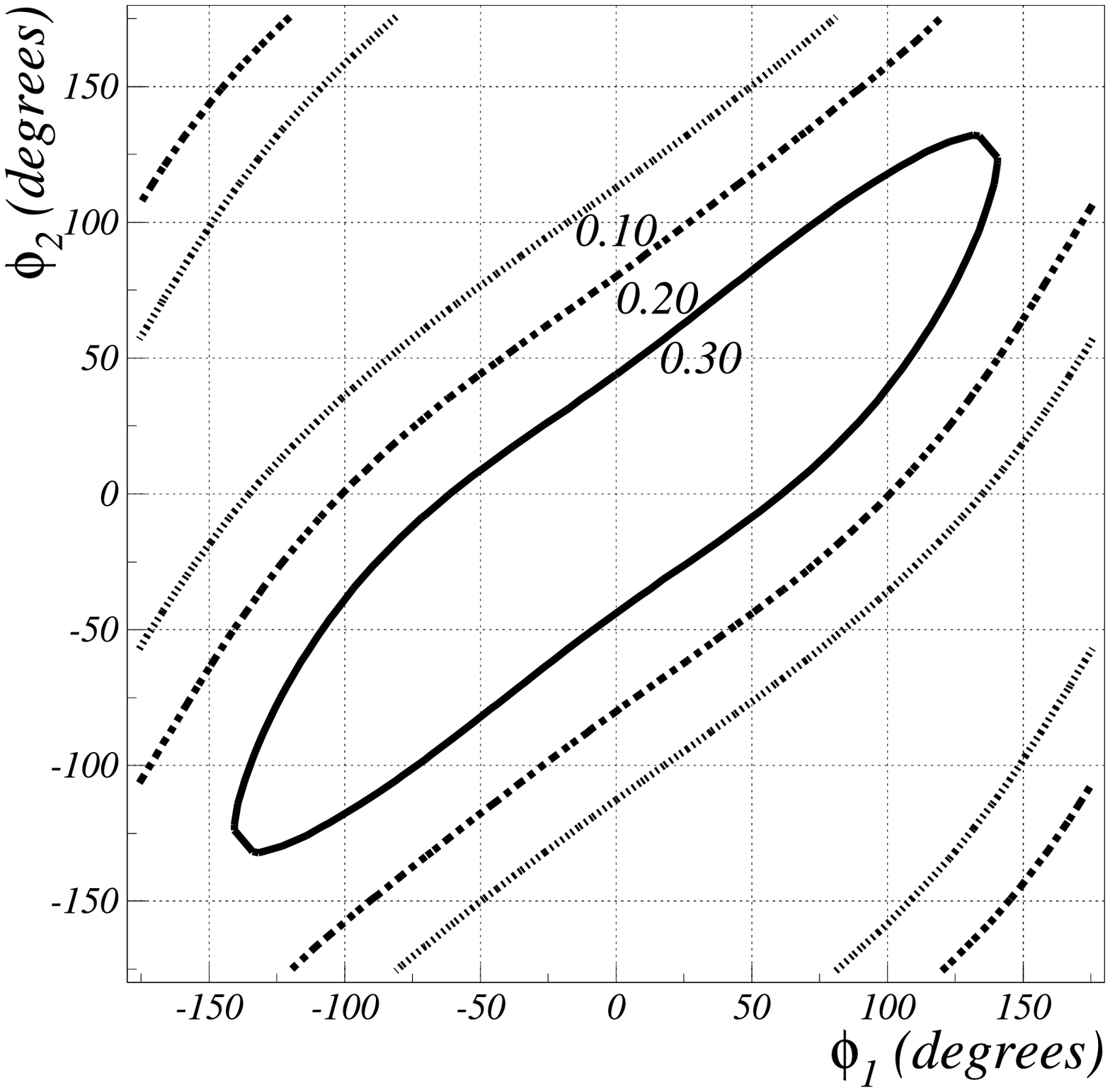}
\includegraphics[origin=c, angle=0, scale=0.46]{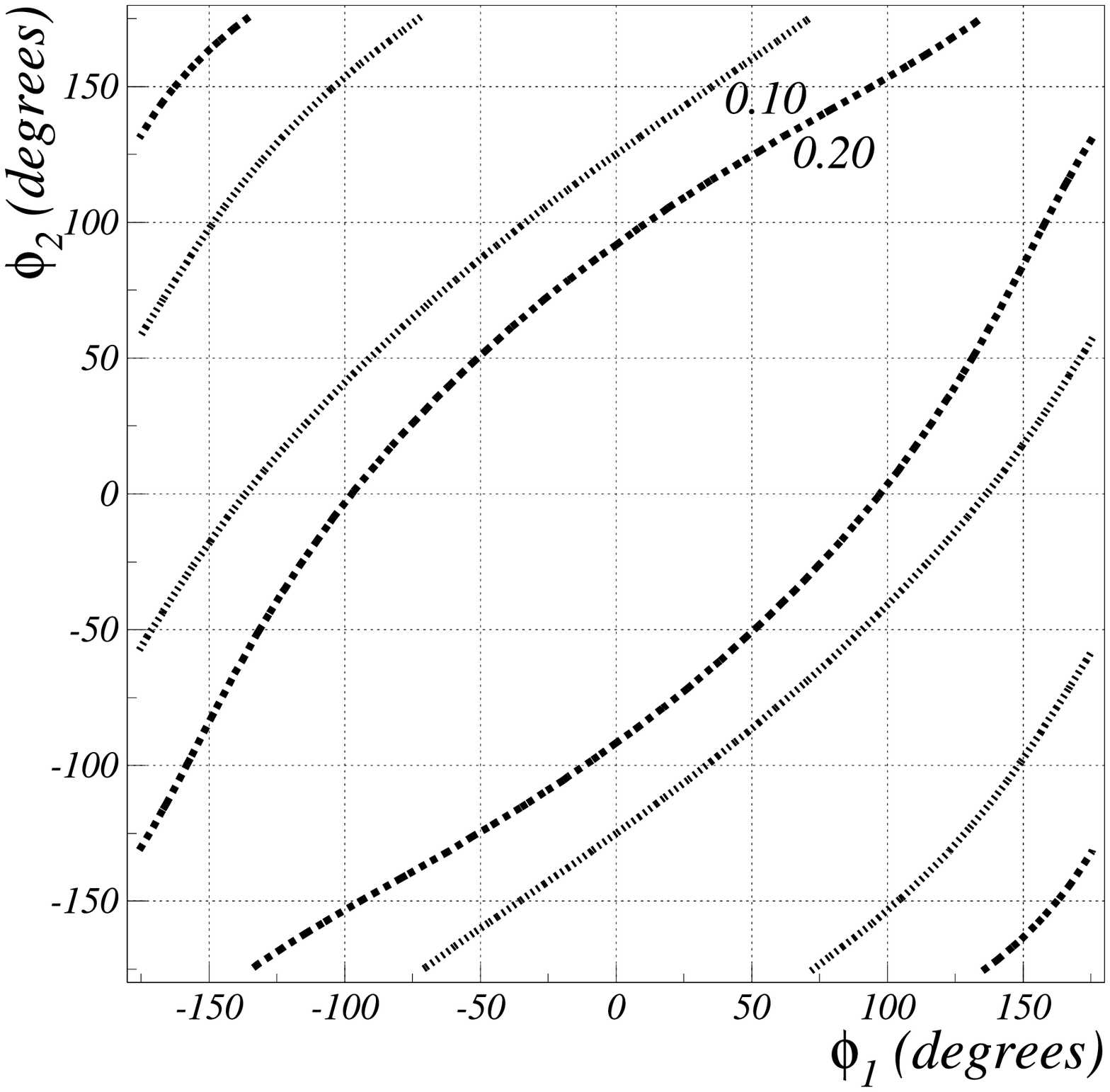}
\caption{Contours of $\hat\sigma_{\mu\mu e}$ (left), $\hat\sigma_{\mu\mu \mu}$ (right) in the $\phi_1$-$\phi_2$ plane for $m_0=0.2$ eV (quasi-degenerate neutrinos). Other parameters are fixed as in Fig.~\ref{fig:m0}.}
\label{fig:majorana3}
\end{center}
\end{figure}

We now study the effect of the Majorana phases $\phi_{1,2}$ on $\hat\sigma_{\ell\ell\ell}$.  It is known that such phases greatly affect ${\cal B}_{\ell\ell}$ \cite{Chun:2003ej,Garayoa:2007fw,Perez:2008ha}, although they have no effect on ${\cal B}_{\ell\nu}$ \cite{Perez:2008ha}.  It is evident from Fig.~\ref{fig:m0} and Fig.~\ref{fig:lhcm0} that the four dominant channels have a similar magnitude as $m_0$ approaches $0.2$ eV (and beyond).  In Figs.~\ref{fig:majorana1}, \ref{fig:majorana2}, and \ref{fig:majorana3}, the effect of the Majorana phases on $\hat\sigma_{\ell\ell\ell}$ is studied for $m_0=0.2$ eV at the Tevatron.  In Fig.~\ref{fig:majorana1} for the $eee$ and $ee\mu$ channels, the rates are higher for smaller $\phi_1$.  Notably, there is no dependence on $\phi_2$ for $eee$, and only a small dependence on $\phi_2$ for $ee\mu$,
the reason being our choice of $\sin^22\theta_{13}=0$.
In Fig.~\ref{fig:majorana2} for the $e\mu e$ and $e\mu\mu$ channels, the rates are higher for larger $\phi_1$ and observable values are attainable (in contrast to Fig.~\ref{fig:m0} and Fig.~\ref{fig:lhcm0} where the rates were negligible).  Again, there is no dependence on $\phi_2$ for $e\mu e$ and only a minor dependence for $e\mu\mu$.  In Fig.~\ref{fig:majorana3} for the $\mu\mu e$ and $\mu\mu\mu$ channels, both $\phi_1$ and $\phi_2$ have a large effect on the rates, with the largest values occurring near the $\phi_1 = \phi_2$ line.  It is noted that there is the symmetry $\phi_{1,2} \leftrightarrow -\phi_{1,2}$  in all the channels.

In Tables~\ref{normal_numbers} (for the normal neutrino mass hierarchy) and \ref{inverted_numbers} (for the inverted neutrino mass hierarchy) the approximate allowed ranges of $\hat\sigma_{\ell\ell\ell}$ are shown for five different values of $m_0$, for the Tevatron.  The Majorana phases, which have the dominant effect on BR$(H^{\pm\pm}\to \ell^\pm\ell^\pm)$, are varied $-\pi < \phi_1,\phi_2 < \pi$, and the other parameters are fixed as in Fig.~\ref{fig:m0}. In Tables \ref{normal_numbers} and \ref{inverted_numbers}, the first number and second number refer to the minimum and maximum values of $\hat\sigma_{\ell\ell\ell}$, respectively, and ``0.0'' signifies $\hat\sigma_{\ell\ell\ell} < 0.01$.

In order to show what value of $\hat\sigma_{\ell\ell\ell}$ could give an observable rate at the Tevatron and the LHC, we give in Table~\ref{Event-numbers} the number of three-lepton events for the case of $\hat\sigma_{\ell\ell\ell}=1$ ({\it i.e.}, the cross section is the same as that for $pp\to H^{++}H^{--}$).  We take two representative values of $M_{H^{\pm\pm}}$ and realistic luminosities for the Tevatron (${\cal L}=10$ fb$^{-1}$) and LHC (${\cal L}=10$ fb$^{-1}$ and $100$ fb$^{-1}$).  Table~\ref{Event-numbers} suggests that with the same doubly charged Higgs mass and luminosity, the LHC should produce roughly 10 times more 3$\ell$ events than the Tevatron.  It is also clear that $\hat\sigma_{\ell\ell\ell} > 0.1$ would give a sizeable number of events at the LHC if $M_{H^{\pm\pm}}< 250$ GeV, and thus multiple three-lepton signals would be possible.

\begin{table}
\begin{center}
\begin{tabular}{|c|c|c|c|c|c|c|}
\hline
$m_0$  & $e^\pm e^\pm e^\mp$ & $e^\pm e^\pm \mu^\mp$ &  $e^\pm \mu^\pm e^\mp$ &  $e^\pm \mu^\pm\mu^\mp$ & $\mu^\pm \mu^\pm e^\mp$  & $\mu^\pm \mu^\pm \mu^\mp$
\tabularnewline
\hline
0.20 eV&  $0.10/0.24$ & $0.10/0.57$   & $0.0/0.35$ & $0.0/0.38$  & $0.0/0.36$ 
& $0.0/0.25$
\tabularnewline
\hline
0.10 eV  &  $0.08/0.21$ & $0.08/0.56$ & $0.0/0.31$ & $0.0/0.37$  & $0.0/0.35$ 
& $0.0/0.27$ \tabularnewline
\hline
0.05 eV  & $0.06/0.18$ & $0.07/0.49$ & $0.0/0.20$ & $0.0/0.33$  & $0.0/0.31$ 
& $0.02/0.31$            \tabularnewline
\hline
0.01 eV  & $0.0/0.0$ & $0.0/0.08$ & $0.0/0.0$ & $0.0/0.08$  & $0.02/0.07$ 
& $0.28/0.50$
\tabularnewline
\hline 
0 eV & $0.0/0.0$ & $0.0/0.0$ & $0.0/0.0$ & $0.0/0.01$  & $0.0/0.01$ 
& $0.37/0.50$
\tabularnewline
\hline
\end{tabular}
\end{center}
\caption{Approximate permitted ranges of the magnitude of $\hat\sigma_{\ell\ell\ell}$, obtained by varying $-\pi < \phi_1,\phi_2 < \pi$ for several values of $m_0$. Other parameters are fixed as in Fig.~\ref{fig:m0}.  We take the normal neutrino mass hierarchy and the energy of the Tevatron.}
\label{normal_numbers}
\end{table}

\begin{table}
\begin{center}
\begin{tabular}{|c|c|c|c|c|c|c|}
\hline
$m_0$  & $e^\pm e^\pm e^\mp$ & $e^\pm e^\pm \mu^\mp$ &  $e^\pm \mu^\pm e^\mp$ &  $e^\pm \mu^\pm\mu^\mp$ & $\mu^\pm \mu^\pm e^\mp$  & $\mu^\pm \mu^\pm \mu^\mp$
\tabularnewline
\hline
0.20 eV&  $0.10/0.24$ & $0.10/0.57$   & $0.0/0.35$ & $0.0/0.38$  & $0.0/0.36$ 
& $0.0/0.25$
\tabularnewline
\hline
0.10 eV  &  $0.11/0.28$ & $0.10/0.60$ & $0.0/0.40$ &  $0.0/0.38$ &  $0.0/0.37$ & $0.0/0.23$ \tabularnewline
\hline
0.05 eV  &   $0.14/0.35$ &  $0.10/0.61$ & $0.0/0.52$ & $0.0/0.40$ & $0.0/0.38$ & $0.0/0.20$ \tabularnewline
\hline
0.01 eV  & $0.22/0.53$ &  $0.11/0.57$ &   $0.0/0.76$ &  $0.0/0.37$ & $0.01/0.28$ & $0.01/0.14$
\tabularnewline
\hline 
0 eV & $0.22/0.55$ &  $0.12/0.52$ &   $0.0/0.80$ &  $0.0/0.34$ &  $0.06/0.19$ &  $0.03/0.11$ 
\tabularnewline
\hline
\end{tabular}
\end{center}
\caption{Same as Table \ref{normal_numbers} but for the inverted neutrino mass hierarchy.}
\label{inverted_numbers}
\end{table}

\begin{table}
\begin{center}
\begin{tabular}{|c|c|c|c|c|}
\hline
  & ${\cal L}$ (fb$^{-1}$) & $M_{H^{\pm\pm}}$  & $\hat\sigma_{\ell\ell\ell}$ & $N_{\ell\ell\ell}$ \tabularnewline
\hline
Tevatron & 10 & 150 GeV & $\sim 20$ fb & $\sim 200$ \tabularnewline
\hline
LHC & 10 & 150 GeV & $\sim 200$ fb & $\sim 2000$ \tabularnewline
 \hline
LHC & 100 & 250 GeV & $\sim 30$ fb & $\sim 3000$ \tabularnewline
\hline
\end{tabular}
\end{center}
\caption{Number of three-lepton events ($N_{\ell\ell\ell}$) for the case of $\hat\sigma_{\ell\ell\ell}=1$, for representative values of $M_{H^{\pm\pm}}$ and the integrated luminosity (${\cal L}$) of the Tevatron and the LHC.}
\label{Event-numbers}
\end{table}

\section{Summary}

The most recent search for doubly charged Higgs bosons at the Fermilab Tevatron requires three muons in order to reduce the Standard Model backgrounds to an acceptable level.  In these three-lepton ($\ell\ell\ell$) exclusive channels the greatest detection potential is for $\ell=e$ and $\mu$, for which there are six distinct channels.  In the Higgs Triplet Model the magnitude of their cross sections is determined by the parameters of the neutrino mass matrix.  We analyze the pattern of the cross sections and show that any of the six channels could be dominant.  We also show that their rates are significantly enhanced by the contribution of the subprocess $q \overline{q'} \to H^{\pm\pm} H^{\mp}$, followed by the decay $H^\pm\to \ell^\pm\nu_\ell$.  The discovery potential for $H^{\pm\pm}$ at the Tevatron will remain competitive even into the era of the LHC, and we encourage searches for $H^{\pm\pm}$ in the above six exclusive channels.

{\it Acknowledgments}: We thank T.~J.~Kim and J.~A.~Aguilar-Saavedra for useful communications.  This work was financially supported in part by the National Science Council of Taiwan, R.~O.~C.\ under Grant No.~NSC~97-2112-M-008-002-MY3.  A.G.A is supported by the ``National Central University Plan to Develop First-class Universities and Top-level Research Centers''.


\begin{thebibliography}{99}

\bibitem{Fukuda:1998mi}
  Y.~Fukuda {\it et al.}  [Super-Kamiokande Collaboration],
  Phys.\ Rev.\ Lett.\  {\bf 81}, 1562 (1998).

\bibitem{Kuno:1999jp}
  Y.~Kuno and Y.~Okada,
  Rev.\ Mod.\ Phys.\  {\bf 73}, 151 (2001).

\bibitem{Minkowski:1977sc}
  P.~Minkowski,
  Phys.\ Lett.\ B {\bf 67} (1977) 421;
  T. Yanagida, in {\it Proceedings of the Workshop on the Unified Theory
   and the Baryon Number in the Universe}, eds. O. Sawada et al., (KEK
   Report~79-18, Tsukuba, 1979), p.~95;
  M. Gell-Mann, P. Ramond and R. Slansky,
   in {\it Supergravity}, eds. P. van Nieuwenhuizen et al.,
   (North-Holland, 1979), p.~315;
   R.~N.~Mohapatra and G.~Senjanovi\'c,
  Phys.\ Rev.\ Lett.\  {\bf 44} (1980) 912.


\bibitem{Zee:1980ai}
  A.~Zee,
  Phys.\ Lett.\  B {\bf 93}, 389 (1980)
  [Erratum-ibid.\  B {\bf 95}, 461 (1980)];
  A.~Zee,
  Nucl.\ Phys.\  B {\bf 264}, 99 (1986);
  K.~S.~Babu,
  Phys.\ Lett.\  B {\bf 203}, 132 (1988).

\bibitem{Konetschny:1977bn}
  W.~Konetschny and W.~Kummer,
  Phys.\ Lett.\  B {\bf 70}, 433 (1977);
  M.~Magg and C.~Wetterich,
  Phys.\ Lett.\  B {\bf 94}, 61 (1980);
  T.~P.~Cheng and L.~F.~Li,
  Phys.\ Rev.\  D {\bf 22}, 2860 (1980).

\bibitem{Schechter:1980gr}
J.~Schechter and J.~W.~F.~Valle,
Phys.\ Rev.\ D {\bf 22}, 2227 (1980);


\bibitem{Mohapatra:1980yp}
  R.~N.~Mohapatra and G.~Senjanovic,
  Phys.\ Rev.\  D {\bf 23}, 165 (1981);
  G.~Lazarides, Q.~Shafi and C.~Wetterich,
  Nucl.\ Phys.\  B {\bf 181}, 287 (1981).


\bibitem{Gunion:1989in}
  J.~F.~Gunion, J.~Grifols, A.~Mendez, B.~Kayser and F.~I.~Olness,
  Phys.\ Rev.\ D {\bf 40}, 1546 (1989);
  J.~F.~Gunion, C.~Loomis and K.~T.~Pitts,
  eConf {\bf C960625}, LTH096 (1996)
  [arXiv:hep-ph/9610237];
  K.~Huitu, J.~Maalampi, A.~Pietila and M.~Raidal,
  Nucl.\ Phys.\ B {\bf 487}, 27 (1997);
  M.~Muhlleitner and M.~Spira,
  Phys.\ Rev.\  D {\bf 68}, 117701 (2003).




\bibitem{Han:2007bk}
  T.~Han, B.~Mukhopadhyaya, Z.~Si and K.~Wang,
  Phys.\ Rev.\  D {\bf 76}, 075013 (2007).

\bibitem{Dion:1998pw}
  B.~Dion, T.~Gregoire, D.~London, L.~Marleau and H.~Nadeau,
  Phys.\ Rev.\  D {\bf 59}, 075006 (1999).

\bibitem{Akeroyd:2005gt}
  A.~G.~Akeroyd and M.~Aoki,
  Phys.\ Rev.\ D {\bf 72}, 035011 (2005).

\bibitem{Acosta:2004uj}
  D.~E.~Acosta {\it et al.}  [CDF Collaboration],
  Phys.\ Rev.\ Lett.\  {\bf 93}, 221802 (2004).
\bibitem{Abazov:2004au}
  V.~M.~Abazov {\it et al.}  [D0 Collaboration],
  Phys.\ Rev.\ Lett.\  {\bf 93}, 141801 (2004).
\bibitem{:2008iy}
  V.~M.~Abazov {\it et al.}  [D0 Collaboration],
  Phys.\ Rev.\ Lett.\  {\bf 101}, 071803 (2008).
\bibitem{Aaltonen:2008ip}
  T.~Aaltonen {\it et al.}  [The CDF Collaboration],
  Phys.\ Rev.\ Lett.\  {\bf 101}, 121801 (2008).



\bibitem{Ma:2000wp}
  E.~Ma, M.~Raidal and U.~Sarkar,
  Phys.\ Rev.\ Lett.\  {\bf 85}, 3769 (2000);
  E.~Ma, M.~Raidal and U.~Sarkar,
  Nucl.\ Phys.\ B {\bf 615}, 313 (2001).

\bibitem{Pontecorvo:1957qd}
  B.~Pontecorvo,
  Sov.\ Phys.\ JETP {\bf 7}, 172 (1958)
  [Zh.\ Eksp.\ Teor.\ Fiz.\  {\bf 34}, 247 (1957)];
  Z.~Maki, M.~Nakagawa and S.~Sakata,
  Prog.\ Theor.\ Phys.\  {\bf 28}, 870 (1962).

\bibitem{Bilenky:1980cx}
  S.~M.~Bilenky, J.~Hosek and S.~T.~Petcov,
  Phys.\ Lett.\  B {\bf 94}, 495 (1980);
  M.~Doi, T.~Kotani, H.~Nishiura, K.~Okuda and E.~Takasugi,
  Phys.\ Lett.\  B {\bf 102}, 323 (1981).

\bibitem{Chakrabarti:1998qy}
  S.~Chakrabarti, D.~Choudhury, R.~M.~Godbole and B.~Mukhopadhyaya,
  Phys.\ Lett.\  B {\bf 434}, 347 (1998).

\bibitem{Perez:2008ha}
  P.~Fileviez Perez, T.~Han, G.~y.~Huang, T.~Li and K.~Wang,
  Phys.\ Rev.\  D {\bf 78}, 015018 (2008).

\bibitem{Chun:2003ej}
  E.~J.~Chun, K.~Y.~Lee and S.~C.~Park,
  Phys.\ Lett.\  B {\bf 566}, 142 (2003).

\bibitem{Garayoa:2007fw}
  J.~Garayoa and T.~Schwetz,
  JHEP {\bf 0803}, 009 (2008);
  A.~G.~Akeroyd, M.~Aoki and H.~Sugiyama,
  Phys.\ Rev.\  D {\bf 77}, 075010 (2008);
  M.~Kadastik, M.~Raidal and L.~Rebane,
  Phys.\ Rev.\  D {\bf 77}, 115023 (2008).

\bibitem{Cuypers:1996ia}
  F.~Cuypers and S.~Davidson,
  Eur.\ Phys.\ J.\  C {\bf 2}, 503 (1998).


\bibitem{Kakizaki:2003jk}
  M.~Kakizaki, Y.~Ogura and F.~Shima,
  Phys.\ Lett.\  B {\bf 566}, 210 (2003).

\bibitem{Akeroyd:2009nu}
  A.~G.~Akeroyd, M.~Aoki and H.~Sugiyama,
  Phys.\ Rev.\  D {\bf 79}, 113010 (2009).


\bibitem{Abbiendi:2001cr}
  G.~Abbiendi {\it et al.}  [OPAL Collaboration],
  Phys.\ Lett.\  B {\bf 526}, 221 (2002);
  P.~Achard {\it et al.}  [L3 Collaboration],
  Phys.\ Lett.\  B {\bf 576}, 18 (2003);
  J.~Abdallah {\it et al.}  [DELPHI Collaboration],
  Phys.\ Lett.\  B {\bf 552}, 127 (2003).

\bibitem{Aktas:2006nu}
  A.~Aktas {\it et al.}  [H1 Collaboration],
  Phys.\ Lett.\  B {\bf 638}, 432 (2006).


\bibitem{Huitu:1996su}
  K.~Huitu, J.~Maalampi, A.~Pietila and M.~Raidal,
  Nucl.\ Phys.\ B {\bf 487}, 27 (1997);
  J.~Maalampi and N.~Romanenko,
  Phys.\ Lett.\ B {\bf 532}, 202 (2002).

\bibitem{delAguila:2008cj}
  F.~del Aguila and J.~A.~Aguilar-Saavedra,
  Nucl.\ Phys.\  B {\bf 813}, 22 (2009).

\bibitem{Hektor:2007uu}
  A.~Hektor, M.~Kadastik, M.~Muntel, M.~Raidal and L.~Rebane,
  Nucl.\ Phys.\  B {\bf 787}, 198 (2007);
  T.~Rommerskirchen and T.~Hebbeker,
  J.\ Phys.\ G {\bf 33}, N47 (2007);
  G.~Azuelos, K.~Benslama and J.~Ferland,
  J.\ Phys.\ G {\bf 32}, 73 (2006).



\end{thebibliography}
\end{document}